\documentclass{revtex4}
\usepackage{amsfonts}
\usepackage{amsmath}
\usepackage{eurosym}
\usepackage{hyperref}
\usepackage{graphicx}
\begin{document}

\title{On the Klein-Gordon scalar field oscillators in a spacetime with spiral-like dislocations in external magnetic fields}
\author{Omar Mustafa}
\email{omar.mustafa@emu.edu.tr (Corr. Auth.)}
\affiliation{Department of Physics, Eastern Mediterranean University, 99628, G. Magusa,
north Cyprus, Mersin 10 - T\"{u}rkiye.}
\author{Abdullah Guvendi}
\email{abdullah.guvendi@erzurum.edu.tr}
\affiliation{Department of Basic Sciences, Erzurum Technical University, 25050, Erzurum, T\"{u}rkiye.}

\begin{abstract}
We investigate the effects of two types of spiral dislocation (the distortion of the radial line, labeled as spiral dislocation I, and the distortion of a circle, labeled as spiral dislocation II) on the relativistic dynamics of the Klein-Gordon (KG) oscillator fields, both in the presence and absence of external magnetic fields. In this context, our investigations show that while spiral dislocation I affects the energies of the KG oscillators (with or without the magnetic field), spiral dislocation II has, interestingly, no effect on the KG oscillator's energies unless a magnetic field is applied. However, for both types of spiral dislocations, we observe that the corresponding wave functions incorporate the effects of the dislocation parameter. Our findings are based on the exact solvability and conditional exact solvability (associated with the biconfluent Heun polynomials) of the KG oscillators (with or without the magnetic field, respectively) for spiral dislocation I, and the exact solvability of the KG oscillators (with or without the magnetic field) for spiral dislocation II. The exact solvability of the latter suggests that the oscillator's frequency is solely determined by the magnetic field strength. \\

\textbf{PACS }numbers\textbf{: }03.65.Ge,03.65.Pm,02.40.Gh\\

\textbf{Keywords:} Klein-Gordon (KG) oscillator; Spacetime with spiral dislocations; Magnetic fields; Special functions; Conditional exact solvability of the biconfluent Heun polynomials.
\end{abstract}

\maketitle

\section{Introduction}

Topological defects in the spacetime fabric are predictions of the grand
unified theory \cite{1.1,1.2,1.3,1.4}. Such defects are observed to modify
the dynamics as well as the spectroscopic structure of the quantum
particles, e.g., \cite{1.5,1.6,1.7,1.8,1.9,1.10}. Topological defects,
moreover, have stimulated a vast number of investigations in different field
of physics \cite{1.11,1.12,1.121,1.13,1.14}. Cosmic strings \cite{1.15,1.16},
global monopoles \cite{1.17}, and domain walls \cite{1.2,1.3}, are examples
of topological defects. Distortions, on the other hand, are line-like
defects characterized by a delta-function-valued curvature (classified as
disclination) and torsion (classified as dislocation) distributions that
result in rotational and translational holonomy \cite{1.12}. Dislocations
may, nevertheless, be in the form of a spiral-type or a screw-type \cite%
{1.12,1.121,1.13,1.14,1.18, 1.19,1.20}. Two types of spiral dislocations form the scope
of the current study: a distortion of the radial line into spiral \cite{1.20}
and a distortion of a circle into spiral ( i.e., edge dislocation) \cite%
{1.18,1.20}.

The harmonic oscillator is well known to form a reference and a fundamental
model in\ textbooks' quantum mechanics, relativistic and non-relativistic,
in the flat static Minkowski spacetime. Gravitational force fields, on the
other hands, generated by different curved spacetime fabrics with
topological defects, indulging non-inertial and/or static disclination
and/or dislocation, are shown to introduce intriguing effects on the
dynamics of relativistic and non-relativistic quantum particles.
Analogously, therefore, it should be of a fundamental interest to
investigate the gravitational field effects on the harmonic oscillator in
different spacetime backgrounds and under different topological defects.
Dirac oscillators \cite{1.21}, for example, are studied in cosmic string
spacetime \cite{1.22} and in a G\"{o}del type Som-Raychaudhuri spacetime 
\cite{1.23,1.24}. The Klein-Gordon (KG) oscillators \cite{1.25} are studied
in G\"{o}del-type spacetime \cite{1.26,1.27,1.28,1.29,1.30,1.31,1.32,1.33},
in cosmic string spacetime and Kaluza-Klein theory \cite
{1.32,1.34,1.35,1.36,1.37,1.38}.  To the best of our knowledge, the KG-oscillators in a spacetime with spiral-like dislocations in external magnetic fields have never been investigated in the literature before. A scalar quantum field with non-minimal coupling can model a fermion-antifermion pair with oscillator interactions \cite{guv-epjc}. This system is particularly valuable for studying the dynamics of coupled pairs in condensed matter systems, especially in the presence of external driving forces such as magnetic fields. Consequently, exact solutions to the corresponding wave equations in spacetime with topological defects \cite{book}, or in materials containing spiral-like dislocations, are crucial for understanding the electronic, magnetic, and optical properties of these materials \cite{book,nat}, as well as the impact of spiral dislocations on quantum revivals \cite{bakke}. The current study, we believe, would offer some detailed and interesting insights into how such spiral dislocations and external magnetic fields influence the quantum behavior of relativistic scalar oscillator fields. 

The organization of our study is in order. In section 2, we recollect the relevant and preliminary mathematical background for KG-oscillators in a spacetime with spiral dislocations in external magnetic fields. In section 3, we discuss and report the exact analytical solution for KG-oscillators in a spacetime with spiral dislocation I without a magnetic field. A conditional exact solvability (associated with the biconfluent Heun polynomials) for the KG-oscillators in a spacetime with spiral dislocation I with a magnetic field are discussed and reported in section 4. The KG-oscillators in a spacetime with spiral dislocation II with a magnetic field, discussed and reported in section 5, are found to be exactly analytically solvable. Interestingly, to our surprise, we found that whilst the dislocation parameter leaves its fingerprints on the energies for KG-oscillators in a spacetime with spiral dislocation I (without and with the magnetic field), it leaves the energies for KG-oscillators in a spacetime with spiral dislocation II without any dislocation parameter's trace unless a non-zero magnetic field is applied. We summarize and discuss our findings in section 7.

\section{KG-oscillators in a spacetime with spiral dislocations in external magnetic fields}\label{sec2}

In this section, we consider a spacetime with two types of spiral
dislocations \cite{1.20}. The first of which is a distortion of the radial
line into spiral (to be labeled Spiral dislocation I, hereinafter) with a
line element metric, in $\hbar =c=1$ units, reads 
\begin{equation}
ds^{2}=-dt^{2}+\left( 1+\beta ^{2}r^{2}\right) \,dr^{2}+2\beta
r^{2}drd\varphi +r^{2}d\varphi ^{2}+dz^{2},  \label{2.1}
\end{equation}%
where the covariant and contravariant metric tensors are, respectively,
given by%
\begin{equation}
g_{\mu \nu }=\left( 
\begin{tabular}{cccc}
$-1\smallskip $ & $\,0\,$ & $0$ & $\,0$ \\ 
$0$ & $\left( 1+\beta ^{2}r^{2}\right) \,$ & $\beta r^{2}$ & $0$ \\ 
$0$ & $\beta r^{2}$ & $r^{2}\,$ & $0$ \\ 
$0$ & $0$ & $0 $ & $1$%
\end{tabular}%
\right) \,;\;g^{\mu \nu }=\left( 
\begin{tabular}{cccc}
$-1\smallskip $ & $\,0\,$ & $0$ & $0$ \\ 
$0$ & $\,1\smallskip $ & -$\beta $ & $0$ \\ 
$0$ & $\,-\beta $ & $\left( \,\frac{1}{r^{2}}+\beta \right) \,$ & $0$ \\ 
$0$ & $0$ & $0$ & $1$%
\end{tabular}%
\right) ;\ \det \left( g_{\mu \nu }\right) =g=-r^{2}.  \label{2.2}
\end{equation}%
Whereas, the second is a distortion of a circle into spiral, i.e., edge
dislocation,(to be labeled Spiral dislocation II, hereinafter) with a line
element metric
\begin{equation}
ds^{2}=-dt^{2}+\,dr^{2}+2\beta drd\varphi +\left( \beta ^{2}+r^{2}\right)
d\varphi ^{2}+dz^{2},  \label{2.3}
\end{equation}%
where the corresponding metric tensors, respectively, read%
\begin{equation}
g_{\mu \nu }=\left( 
\begin{tabular}{cccc}
$-1\smallskip $ & $\,0\,$ & $0$ & $\,0$ \\ 
$0$ & $1\,$ & $\beta $ & $0$ \\ 
$0$ & $\beta $ & $\left( \beta ^{2}+r^{2}\right) \,$ & $0$ \\ 
$0$ & $0$ & $0$ & $1$%
\end{tabular}%
\right) \,;\;g^{\mu \nu }=\left( 
\begin{tabular}{cccc}
$-1\smallskip $ & $\,0\,$ & $0$ & $0$ \\ 
$0$ & $\left( \,\frac{\beta ^{2}}{r^{2}}+1\right) $ & $-\frac{\beta }{r^{2}}$
& $0$ \\ 
$0$ & $-\frac{\beta }{r^{2}}$ & $\,\frac{1}{r^{2}}\,$ & $0$ \\ 
$0$ & $0$ & $0\,$ & $\,1$%
\end{tabular}
\right) ;\ \det \left( g_{\mu \nu }\right) =g=-r^{2}.  \label{2.4}
\end{equation}
The covariant KG-equation for a spin-zero scalar particle, of rest mass
energy $m_{\circ }$ (i.e., $m_{\circ }c^{2}$), under the influence of a
minimally coupled electromagnetic vector field $A_{\mu }$, a non-minimally
coupled vector field $\mathcal{F}_{\mu }$, and a Lorentz scalar field $
S\left( r\right) $ is given by
\begin{equation}
\left( \frac{1}{\sqrt{-g}}\tilde{D}_{\mu }^{+}\sqrt{-g}g^{\mu \nu }\tilde{D}
_{\nu }^{-}\right) \,\Psi \left( t,r,\varphi ,z\right) =\left[ m_{\circ
}+S\left( r\right) \right] ^{2}\,\Psi \left( t,r,\varphi ,z\right) ,
\label{2.5}
\end{equation}%
where $\tilde{D}_{\mu }^{\pm }=D_{\mu }\pm \mathcal{F}_{\mu }$ is in a non-minimal coupling form with $\mathcal{F}_{\mu }$ $\in \mathbb{R}$, $D_{\mu }=\partial _{\mu }-ieA_{\mu }$ is the gauge-covariant derivative.
We shall use a vector field $\mathcal{F}_{\mu }=\left( 0,\mathcal{F}
_{r},0,0\right) $. One should notice that $\mathcal{F}_{r}=\eta r$ is used
to incorporate the KG-oscillators in the process.

\section{KG-oscillators in a spacetime with spiral dislocation I}\label{sec2:1}

In this section we consider the KG-oscillator in a spacetime described by
metric (\ref{2.1}). In this case, using (\ref{2.5}) along with the
corresponding contravariant metric tensor components in (\ref{2.2}) we obtain
\begin{gather}
\left\{ \partial _{r}^{2}+\frac{1}{r}\partial _{r}-\mathcal{M}\left(
r\right) -\frac{\beta }{r}\left( \partial _{r}+\mathcal{F}_{r}\right)
\,r\,\left( im-ieA_{\varphi }\right) -\beta \left( im-ieA_{\varphi }\right)
\left( \partial _{r}-\mathcal{F}_{r}\right) \right.  \notag \\
-\left. \left( \frac{1}{r^{2}}+\beta \right) \left( m-eA_{\varphi }\right)
^{2}-2m_{\circ }S\left( r\right) -S\left( r\right) ^{2}+\mathcal{E}\right\}
\psi \left( r\right) =0,  \label{2.6}
\end{gather}
where 
\begin{equation}
\mathcal{E}=E^{2}-\left( m_{\circ }^{2}+k^{2}\right) ,\;\mathcal{M}\left(
r\right) =\mathcal{F}_{r}^{2}+\frac{\mathcal{F}_{r}}{r}+\mathcal{F}
_{r}^{\prime },  \label{2.61}
\end{equation}
\begin{figure}[ht!]  
\centering
\includegraphics[width=0.3\textwidth]{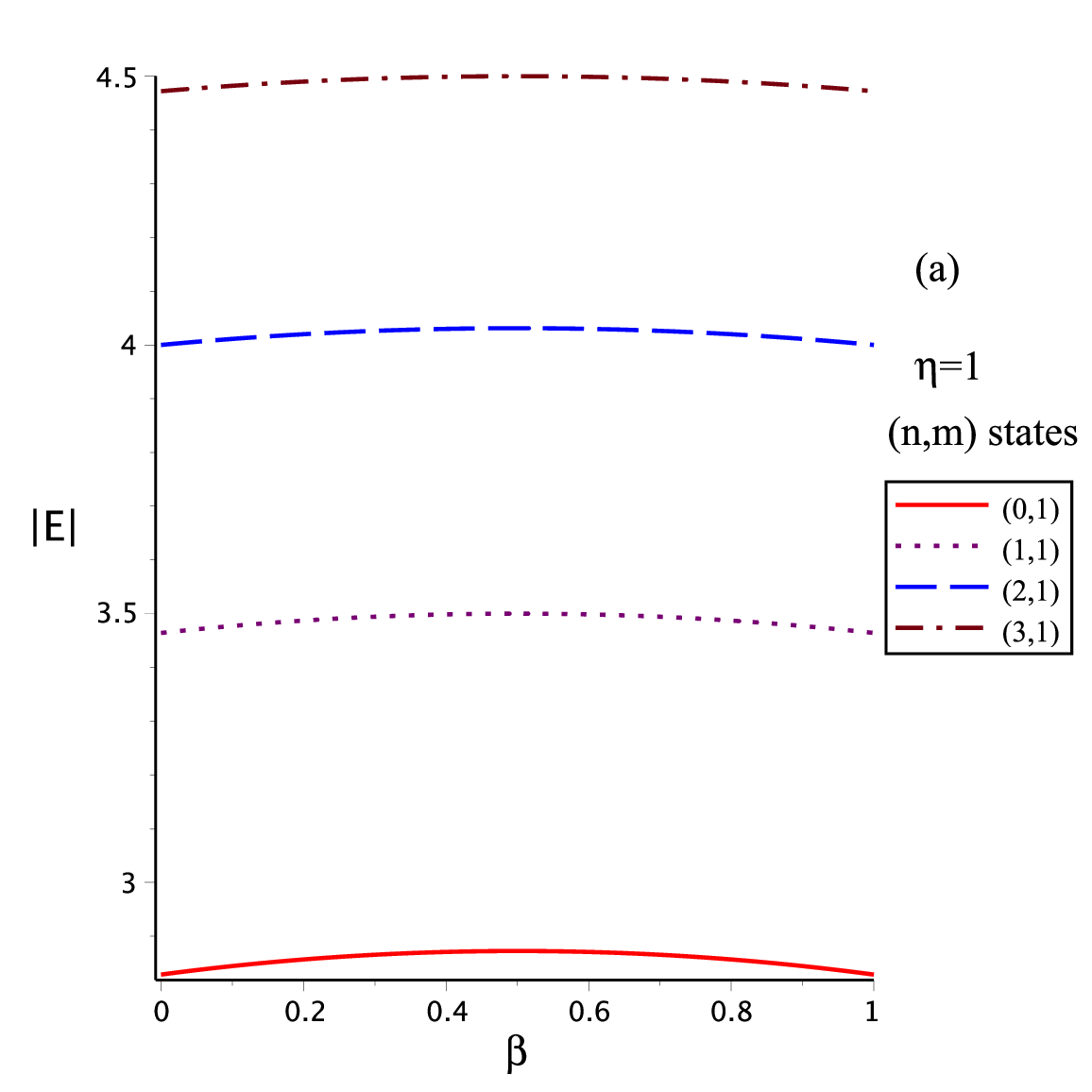}
\includegraphics[width=0.3\textwidth]{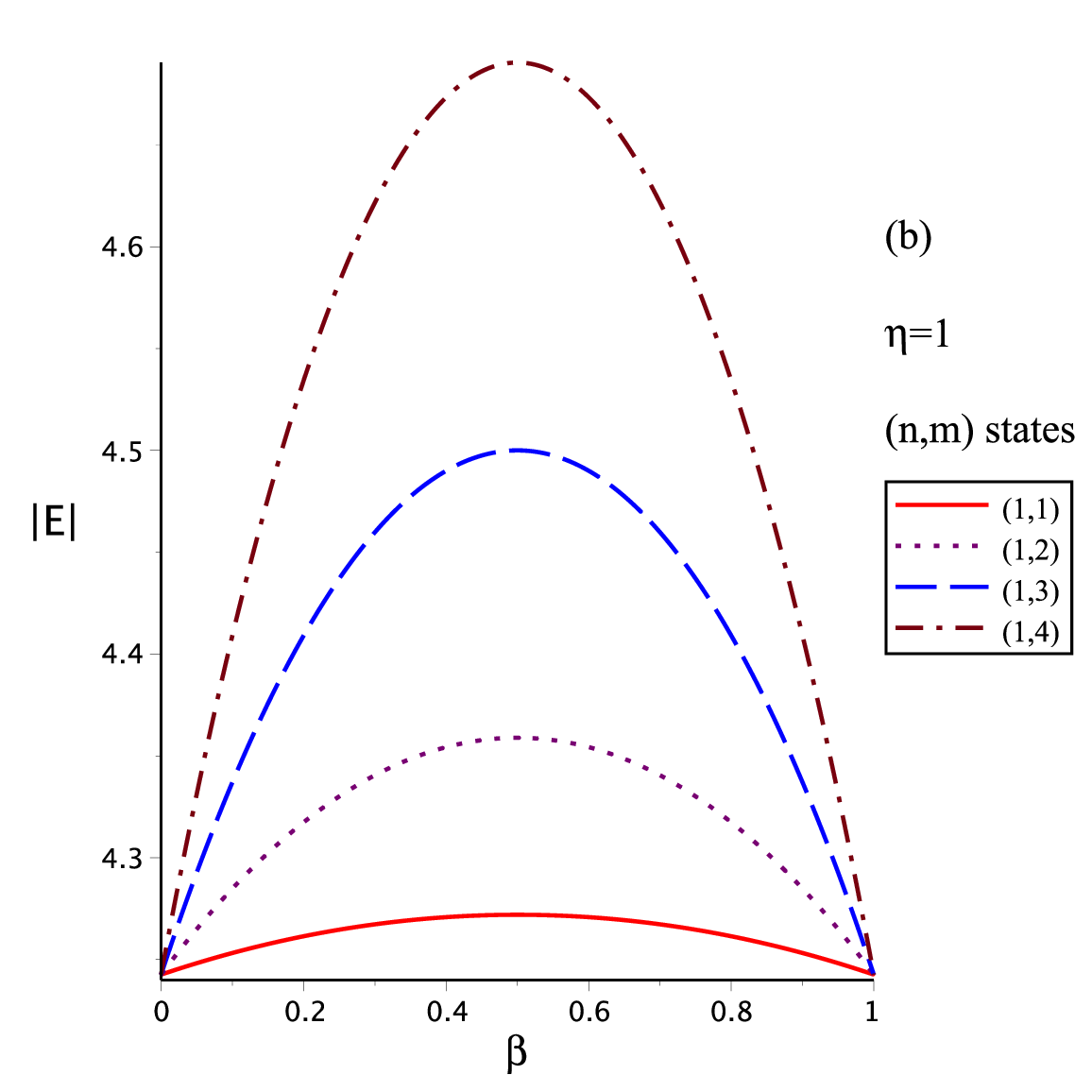}
\includegraphics[width=0.3\textwidth]{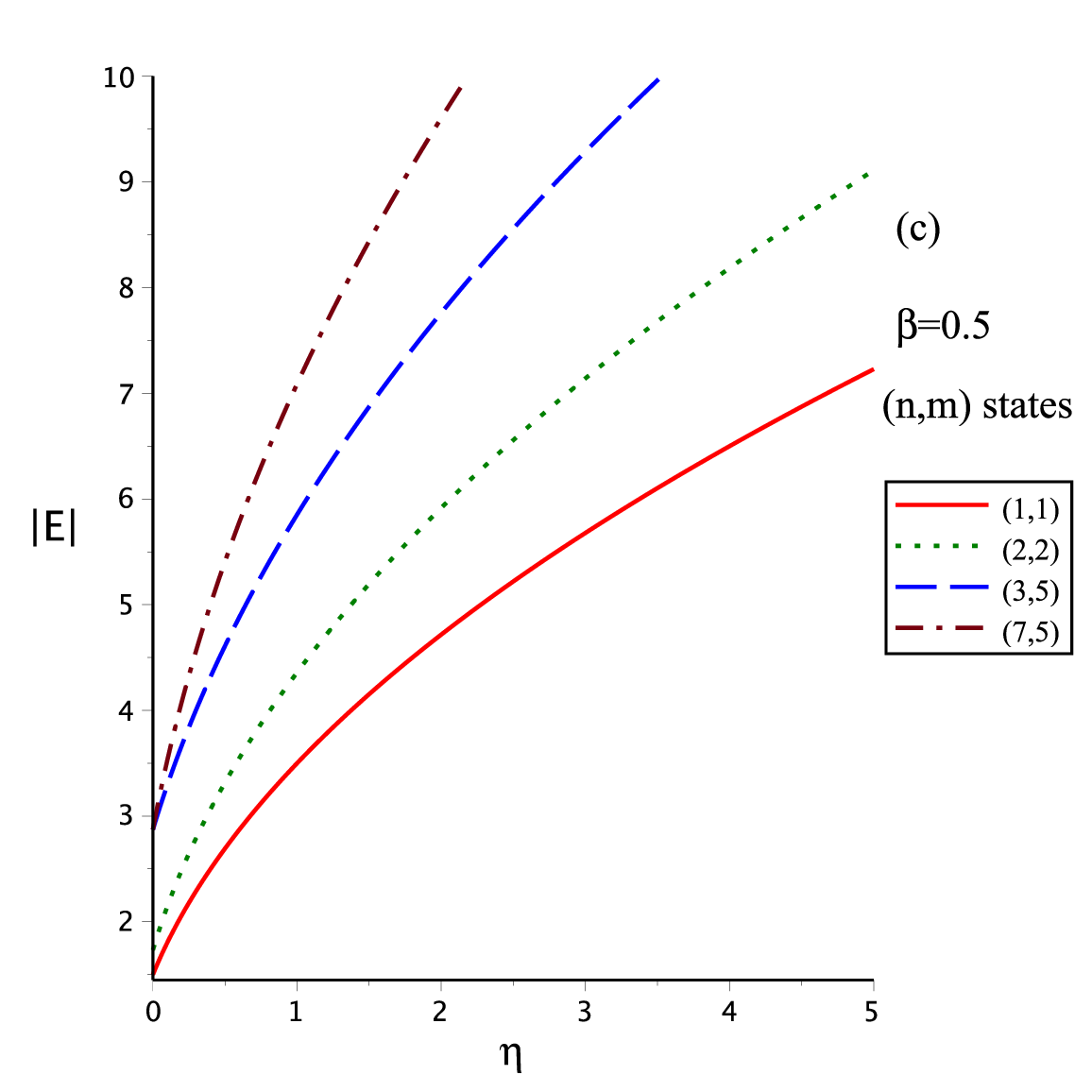}
\caption{\small 
{ The energy levels of Eq. (\ref{2.10}), for the KG-oscillators in a spacetime with spiral dislocation I, where $k=m_{\circ}=e=1$ so that we plot in (a)  $|E|$ against $\beta$, (b) $|E|$ against $\beta$, and (c) $|E|$ against $\eta$.}}
\label{fig1}
\end{figure}%
and we have used $\Psi \left( t,r,\varphi ,z\right) =\exp \left( i\left[
m\varphi +k\,z-Et\right] \right) \psi \left( r\right) $. We now consider $
A_{\varphi }=0=S\left( r\right) $ and $\mathcal{F}_{r}=\eta r$ so that (\ref
{2.6}) now reads%
\begin{equation}
\psi ^{\prime \prime }\left( r\right) +\left( \frac{1}{r}-2i\beta m\right)
\psi ^{\prime }\left( r\right) +\left( \mathcal{\tilde{E}}-\eta ^{2}r^{2}-
\frac{im\beta }{r}-\frac{m^{2}}{r^{2}}\right) \psi \left( r\right) =0,
\label{2.7}
\end{equation}%
where $\mathcal{\tilde{E}=}E^{2}-\left( m_{\circ }^{2}+k^{2}+\beta
m^{2}+2\eta \right) $. The exact solution of which is given by
\begin{equation}
\psi \left( r\right) =\mathcal{N}\,\;\,r^{\left\vert m\right\vert }\exp
\left( im\beta r-\frac{\eta }{2}r^{2}\right) \,_{1}F_{1}\left( \frac{1}{2}%
+\nu -\mu ,\,1+2\nu ,z\right) ,  \label{2.8}
\end{equation}
where%
\begin{equation}
\nu =\frac{\left\vert m\right\vert }{2}\,;\;\mu =\frac{\mathcal{\tilde{E}
+\beta }^{2}m^{2}}{4\eta }\,,\text{ }z=\eta \,r^{2}.  \label{2.9}
\end{equation}
The confluent hypergeometric function/series $_{1}F_{1}\left( \frac{1}{2}
+\nu -\mu ,\,1+2\nu ,z\right) $ is known to become a polynomial of order $
n\geq 0$ when the condition $\frac{1}{2}+\nu -\mu =-n$ is satisfied. This
condition would render the radial function $\psi \left( r\right) $ finite
and square integrable. Consequently, $\frac{1}{2}+\nu -\mu =-n$ yields, with $\grave{\beta}=\beta-\beta^2$,
\begin{equation}
E_{nm}=\pm \sqrt{2\eta \left( 2n+\left\vert m\right\vert +2\right)
+k^{2}+m_{\circ }^{2}+\grave{\beta} m^{2}}.  \label{2.10}
\end{equation}
This result notably suggests that levels $E_{nm}=\pm|E_{nm}|$ are symmetric about $E=0$. Moreover, for a given $n$- state all $m=-|m|$ states collapse into $m=+|m|$ states.  

In Figure \ref{fig1}, we have used the dislocation parameter value $0<\beta<1$ so that complex energies in (\ref{2.10}) are avoided. We observe that the maximum energies $|E_{n,m}|_{max}$ are obtained at $\beta=0.5$, which can be verified by the relation $d|E_{n,m}|/d\beta=0$. This is clear in Fig.s 1(a) and 1(b) where the energies are plotted for different values of $\beta$. At  $\beta=0.5$, moreover, we have plotted the energies against the KG-oscillators' frequency $\eta\geq0$ in 1(c). Therein, we observe energy splittings at $\eta=0$ are due to contribution of  the last term in (\ref{2.10}), which is associated with the square of the magnetic quantum number $m$. 
%

\section{KG-oscillators in a spacetime with spiral dislocation I in a
magnetic field}\label{sec2:2}

In this section, we consider KG-oscillators, $\mathcal{F}_{r}=\eta r$, in a
spacetime with spiral dislocation I (\ref{2.1}) in a magnetic field given by 
$eA_{\varphi }=\tilde{B}\,r$, $\tilde{B}=eB$, without a Lorentz scalar
potential, $S\left( r\right) =0$. Under such settings, equation (\ref{2.6})
would result
\begin{equation}
\psi ^{\prime \prime }\left( r\right) +\left[ 2i\tilde{B}\beta \,r+\frac{1}{r
}-2i\beta m\right] \psi ^{\prime }\left( r\right) +\left[ \bar{\lambda}-
\frac{m^{2}}{r^{2}}-\Omega ^{2}r^{2}+A_{1}r-\frac{A_{2}}{r}\right] \psi
\left( r\right) =0,  \label{2.11}
\end{equation}%
where $\Omega ^{2}=\beta \tilde{B}^{2}+\eta ^{2}$, $A_{1}=2\tilde{B}\beta m$
, $A_{2}=m\left( i\beta -2\tilde{B}\right) $, and
\begin{equation}
\bar{\lambda}=E^{2}-\left( k^{2}+m_{\circ }^{2}+\tilde{B}^{2}+2\eta \right)
+\beta \left( 2i\tilde{B}-m^{2}\right) .  \label{2.12}
\end{equation}%
Equation (\ref{2.11}) admits a finite solution in the form of a biconfluent
Heun functions/series $H\left( r\right) =H_{B}\left( \bar{\alpha},\bar{\beta}%
,\bar{\gamma},\bar{\delta},\bar{z}\right) $ so that 
\begin{equation}
\psi \left( r\right) =r^{\left\vert m\right\vert }\exp \left( \frac{\left[
a-i\beta \tilde{B}\right] \,}{2}r^{2}-m\left[ \frac{\grave{\beta}\tilde{B}}{a%
}-i\beta \right] r\right) \,H_{B}\left( \bar{\alpha},\bar{\beta},\bar{\gamma}%
,\bar{\delta},\bar{z}\right) ,  \label{2.13}
\end{equation}%
where $\grave{\beta}=\beta -\beta ^{2}$, $a=\sqrt{\Omega ^{2}-\beta ^{2}%
\tilde{B}^{2}}=\sqrt{\grave{\beta}\tilde{B}^{2}+\eta ^{2}}$, 
\begin{equation}
\bar{\alpha}=2\left\vert m\right\vert \text{,\ }\bar{\beta}=-\frac{2im\grave{%
\beta}\tilde{B}}{(\grave{\beta}\tilde{B}^{2}+\eta ^{2})^{3/4}}\text{, }\bar{\gamma}=\frac{m^{2}\eta ^{2}\grave{%
\beta}}{(\grave{\beta}\tilde{B}^{2}+\eta ^{2})^{3/2}}-\frac{\lambda _{1}}{\left(\grave{\beta}\tilde{B}^{2}+\eta ^{2}\right)^{1/2}}\text{, }\bar{\delta}=\frac{4im\tilde{B%
}}{(\grave{\beta}\tilde{B}^{2}+\eta ^{2})^{1/4}}\text{, }\bar{z}=i\sqrt{a}r.  \label{2.14}
\end{equation}%
and 
\begin{equation}
\lambda _{1}=E^{2}-k^{2}-m_{\circ }^{2}-\tilde{B}^{2}-2\eta .  \label{2.15}
\end{equation}%
At this point, one should notice that when the magnetic field is switched off  $\tilde{B}=0\rightarrow a=-\eta$  so that the wave function in  (\ref{2.13}) vanishes at $r\rightarrow \infty$.  Obviously, moreover, $\tilde{B}=0\rightarrow a=-\eta$  would result that $\bar{\beta}=0$ and $\bar{\delta}=0$ to consequently yield $$\psi \left( r\right) =r^{\left\vert m\right\vert }\exp \left( -\frac{\eta \,}{2}r^{2}+im\beta r\right) \,H_{B}\left( \bar{\alpha},0,\bar{\gamma}%
,0,\bar{z}\right), $$ where  $$H_{B}\left( \bar{\alpha},0,\bar{\gamma},0,\bar{z}\right)=\,  _{1}F_1\left( \frac{1}{2}+\frac{\bar\alpha}{4}-\frac{\bar\gamma}{4},1+\frac{\bar\alpha}{2},\bar{z}^2\right),$$ and, hence, retrieves, upon the truncation condition $$\frac{1}{2}+\frac{\bar\alpha}{4}-\frac{\bar\gamma}{4}=-n\,;\,\, n\geq 0$$ of the confluent hypergeometric series, the result reported in (\ref{2.10}).
\begin{figure}[ht!]  
\centering
\includegraphics[width=0.3\textwidth]{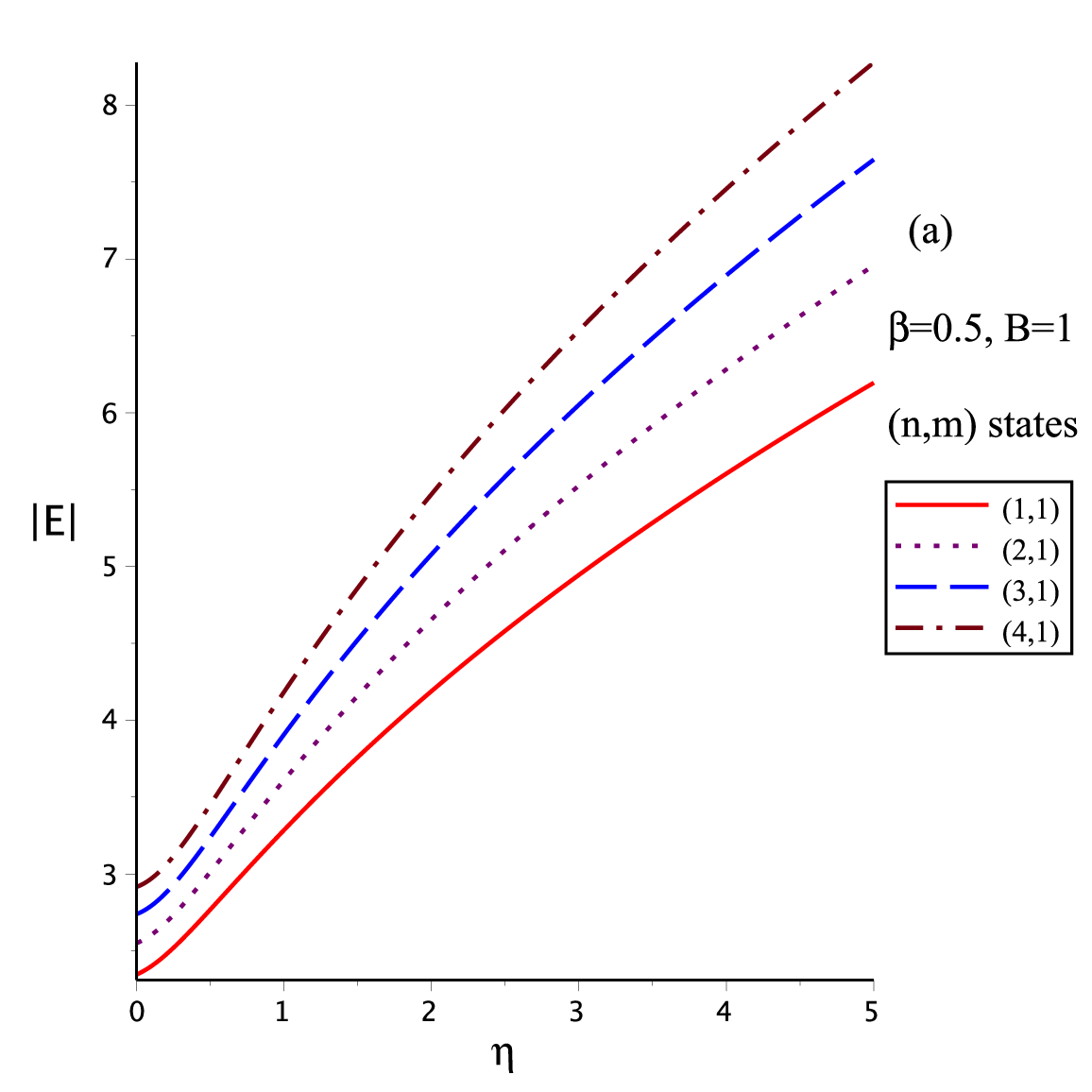}
\includegraphics[width=0.3\textwidth]{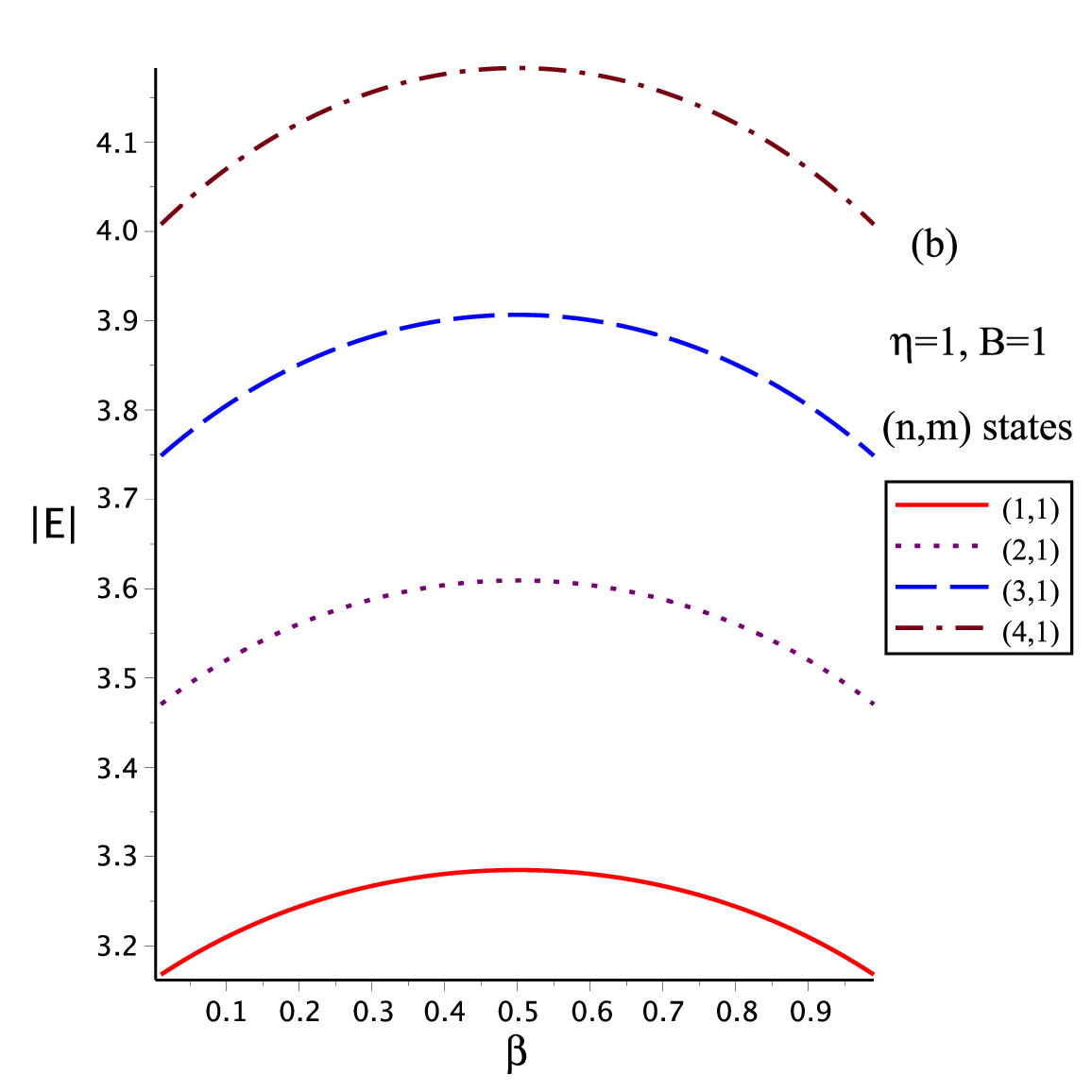}
\includegraphics[width=0.3\textwidth]{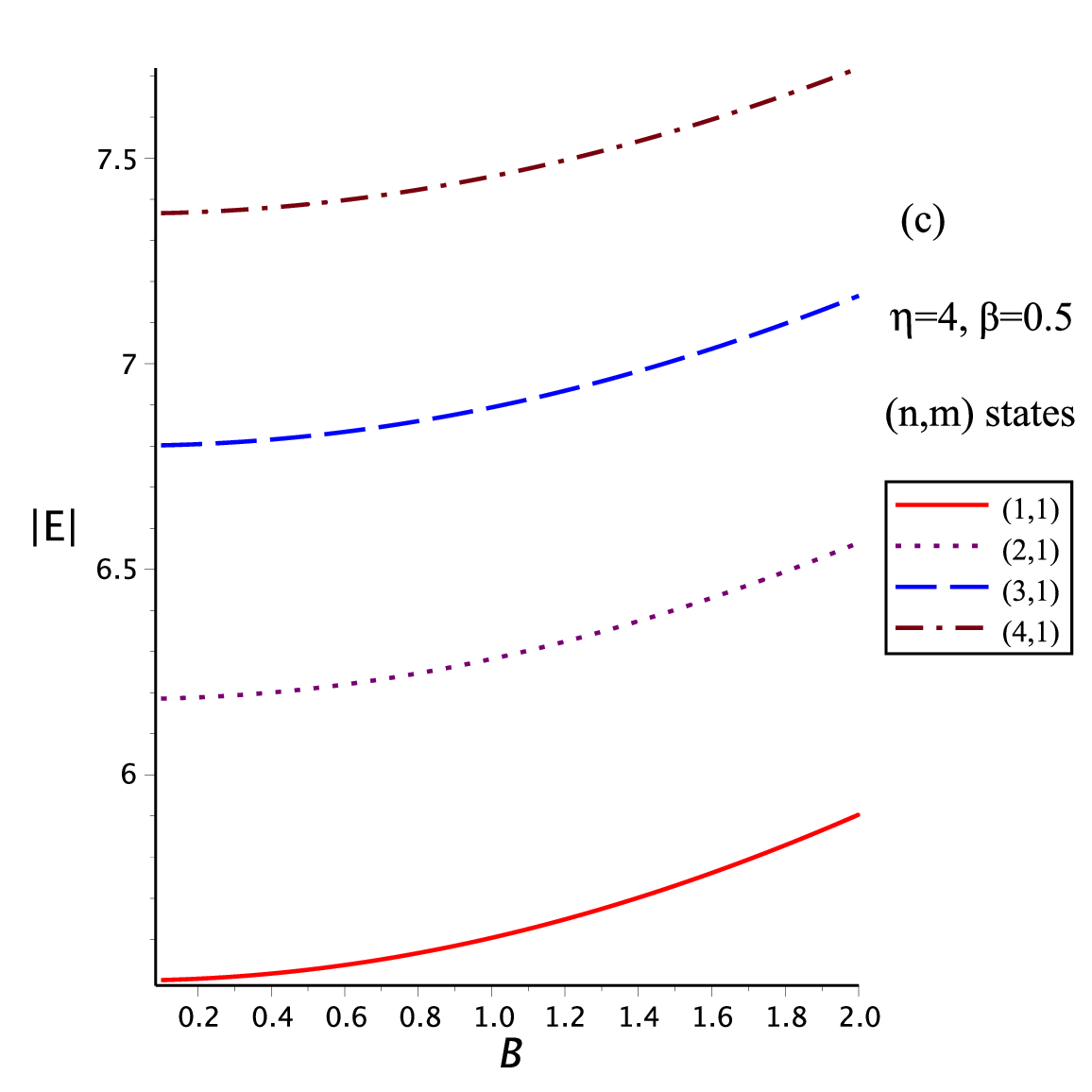}
\caption{\small 
{ The energy levels in Eq. (\ref{2.21}), for the KG-oscillators in a spacetime spiral dislocation I in a magnetic field, with $k=m_{\circ}=e=1$ so that we plot in (a) $|E|$ against $\eta$, (b) $|E|$ against $\beta$, and (c) $|E|$ against the magnetic field $B$.}}
\label{fig2}
\end{figure}%

On the other hand, with the magnetic field switched on, the biconfluent Heun functions/series $H_{B}\left( \bar{\alpha},\bar{\beta},\bar{\gamma},\bar{\delta},z\right) $ are truncated to polynomials of order $%
n\geq 0$ by the conditions that $\bar{\gamma}=2\left( n+1\right) +\bar{\alpha}$  and the series expansion coefficient $A_{n+1}=0$ as described by Ronveaux  \cite{2.1} and Ishkhanyan \cite{2.11}. However, we have very recently introduced truncation conditions that retrieves $\bar{\gamma}=2\left( n+1\right) +\bar{\alpha}$ and, instead
of using $A_{n+1}=0$, we have suggested that $A_{n+1}\neq 0$ but the coefficient associated with it vanish to imply $\bar{\delta}=-\bar{\beta}\left( 2n+\bar{%
\alpha}+3/2\right) $ \cite{2.2,2.3,2.4,2.4.1}. Whilst such conditions would truncate the biconfluent Heun series to polynomials of order $n+1\geq 1$, instead of $n\geq 0$, they would also facilitate conditional exact solvability through some parametric correlation. For more details the reader is advised to see the Appendix of \cite{2.2}. Similar recipe is also successfully implemented for the confluent Heun functions/series as well.
Under the current parametric setting, the two conditions are simplified to read $\bar{\gamma}=\bar{n}+1/2$, and $\bar{\delta}=-\bar{\beta}\bar{n}$,
where $\bar{n}=2n+2\left\vert m\right\vert +3/2$. The use of condition $\bar{%
\delta}=-\bar{\beta}\bar{n}$ would result that
\begin{equation}
2\left(\grave{\beta}\tilde{B}^{2}+\eta ^{2}\right)^{1/2}=\grave{\beta}\bar{n}
\Rightarrow    \grave{\beta}^2 \bar{n}^2-4\grave{\beta}\tilde{B}^{2}-\eta^2=0 \Rightarrow  \grave{\beta}=\frac{1}{\bar{n}^2}\left( 2\tilde{B}^2+\sqrt{4\tilde{B}^2+\eta^2\bar{n}^2}\right)\label{2.17}
\end{equation}
This result suggests that $\grave{\beta}\neq 0\Rightarrow \beta\neq0,1$ otherwise trivial solution is obtained. Yet, $\grave{\beta}>0\Rightarrow 0<\beta<1$ otherwise complex $\bar{n}$ will be obtained which is inconsistent with the fact that our $\bar{n}>3/2$.   
One should observe that $\beta= 0,1$ would yield $a=0$ of (\ref{2.17}) and hence $\bar{\gamma}=\infty$ (i.e., trivial solution is obtained). Such a correlation would, moreover, not only facilitate conditional exact solvability but also identifies the allowed range of the dislocation parameter so that $\beta>1 $, within such conditional solvability of course.  

Next, the condition that $\bar{\gamma}=\bar{n}+1/2$ would yield
\begin{equation}
\frac{m^{2}\eta ^{2}\grave{%
\beta}}{(\grave{\beta}\tilde{B}^{2}+\eta ^{2})^{3/2}}-\frac{\lambda _{1}}{\left(\grave{\beta}\tilde{B}^{2}+\eta ^{2}\right)^{1/2}}=\bar{n}
+1/2 \Rightarrow \lambda _{1}=-{\left(\grave{\beta}\tilde{B}^{2}+\eta ^{2}\right)^{1/2}} \left( \bar{n}+1/2\right)+\frac{m^{2}\eta ^{2}\grave{\beta}}{\grave{\beta}\tilde{B}^{2}+\eta ^{2}}
\label{2.20}
\end{equation}
and eventually imply
\begin{equation}
E_{nm}=\pm \sqrt{-{\left(\grave{\beta}\tilde{B}^{2}+\eta ^{2}\right)^{1/2}} \left( \bar{n}+1/2\right)+\frac{m^{2}\eta ^{2}\grave{\beta}}{\grave{\beta}\tilde{B}^{2}+\eta ^{2}}+k^{2}+m_{\circ }^{2}+\tilde{B}^{2}+2\eta}    \label{2.21}
\end{equation}
It is clear that the energies $E_{nm}$ are symmetric about $E_{nm}=0$. At this point, it is feasibly obvious that we may take $\eta=0$ (i.e., no KG-oscillators) and still obtain energy levels supported by the gravitational field manifestly introduced by the spiral dislocation I. In this case, however, $\sqrt{\grave{\beta}\tilde{B}^2}=-\sqrt{\grave{\beta}}|\tilde{B}|$ so that the radial wave function in (\ref{2.13}) is finite at $r=\infty$. It is wise, therefore, to recast the result in (\ref{2.21}), without any loss of generality,  as
\begin{equation}
E_{nm}=\pm \sqrt{\left|\sqrt{\grave{\beta}\tilde{B}^{2}+\eta ^{2}}\right| \left( \bar{n}+1/2\right)+\frac{m^{2}\eta ^{2}\grave{\beta}}{\grave{\beta}\tilde{B}^{2}+\eta ^{2}}+k^{2}+m_{\circ }^{2}+\tilde{B}^{2}+2\eta}    \label{2.21}
\end{equation}
In Figure \ref{fig2} we plot the KG-oscillators energies $|E_{nm}|$ for different oscillator frequency values in \ref{fig2}(a), for different dislocation parameter $0<\beta<1$ values in \ref{fig2}(b), and for different magnetic field strength $B$ values in \ref{fig2}(c).  We again observe in \ref{fig2}(b) that the $|E_{nm}|_{max}$ is at $\beta=0.5$,  which can be verified by the relation $d|E_{n,m}|/d\beta=0$.  Obviously, moreover, one may observe degeneracies between energy states with the magnetic quantum number $m=\pm|m|$, which are rendered indistinguishable by the form of  (\ref{2.21}) as they are given in either $|m|$ of $\bar{n}$ or $m^2$ of the second term under the square root above.


\section{KG-oscillators in a spacetime with spiral dislocation II in a
magnetic field}\label{sec3:2}

We now consider KG-oscillators in a spacetime with spiral dislocation II given by metric (\ref{2.3}
), along with (\ref{2.4}), and in a magnetic field given by the vector potential $A_{\mu }=\left( 0,0,A_{\varphi },0\right) $,
so that KG-equation (\ref{2.5}) would imply
\begin{gather}
\left\{ \partial _{r}^{2}+\frac{1}{r}\partial _{r}-\mathcal{M}\left(
r\right) +\frac{\beta ^{2}}{r}\left( \partial _{r}+\mathcal{F}_{r}\right) 
\frac{1}{r}\left( \partial _{r}-\mathcal{F}_{r}\right) -\frac{\beta }{r}
\left( \partial _{r}+\mathcal{F}_{r}\right) \,\,\frac{1}{r}\left(
im-ieA_{\varphi }\right) \right.  \notag \\
-\left. \frac{\beta }{r}\left( im-ieA_{\varphi }\right) \left( \partial
_{r}-\mathcal{F}_{r}\right) -\frac{\left( m-eA_{\varphi }\right) ^{2}}{r^{2}}
-2m_{\circ }S\left( r\right) -S\left( r\right) ^{2}+\mathcal{E}\right\} \psi
\left( r\right) =0,  \label{3.1}
\end{gather}
where, again, $\mathcal{E}=E^{2}-\left( m_{\circ }^{2}+k^{2}\right) $ and $%
\Psi \left( t,r,\varphi ,z\right) =\exp \left( i\left[ m\varphi +k\,z-Et
\right] \right) \psi \left( r\right) $ are used here. For the KG-oscillators we again use  $\mathcal{F}_{r}=\eta r$  in the electromagnetic 4-vector potential component  $eA_{\varphi }=\tilde{B}\,r^{2}$ and without the Lorentz scalar potential (i.e., $S\left( r\right) =0$). At this point, we may use the substitution 
\begin{equation}
\psi \left( r\right) =\exp \left( i\left[ \beta ^{2}\tilde{B}+m\right]
\arctan \left( \frac{r}{\beta }\right) -i\beta \tilde{B}r-\frac{\Omega r^2}{2}\right) U\left( r\right) ,  \label{4.1}
\end{equation}
where  $\Omega ^{2}=\eta ^{2}+\tilde{B}^{2}$. This would yield, with the change of variables \ $z=r^{2}$, read
\begin{equation}
4\left( \beta ^{2}+z\right) ^{2}U^{\prime \prime }\left( z\right) -4\left(
\beta ^{2}+z\right) \left( \Omega \beta ^{2}-1+\Omega \,z\right) U^{\prime
}\left( z\right) +\left( a\,z+b\right) U\left( z\right) =0,  \label{4.2}
\end{equation}
where 
\begin{eqnarray}
\tilde{E} &=&E^{2}-k^{2}-m_{\circ }^{2}-2\eta -2\Omega,  \notag \\
a &=&\tilde{E}+2\tilde{B}^{2} \beta ^{2}+2m\tilde{B},  \label{4.3} \\
b &=&\beta ^{4}\tilde{B} ^{2}+ \tilde{E}\beta^2-m^{2}.  \notag
\end{eqnarray}
Equation (\ref{4.2}) admits a solution in the form of confluent
hypergeometric function
\begin{equation}
U(r)=U\left( z(r)\right) =\mathcal{N}\,  (r^2+\beta^2)^{\left( \beta ^{2}
\tilde{B}+\left\vert m\right\vert \right) /2}\,_{1}F_{1}\left( \frac{1}{2}
+\nu -\mu ,1+2\nu ,\Omega (\beta^2+r^2)\right) ,  \label{4.4}
\end{equation}
where
\begin{equation}
\mu =\frac{a+2 \Omega }{
4\Omega },\;\nu =\frac{\beta ^{2}\tilde{B}+\left\vert m\right\vert }{2}.
\label{4.5}
\end{equation}%
\begin{figure}[ht!]  
\centering
\includegraphics[width=0.3\textwidth]{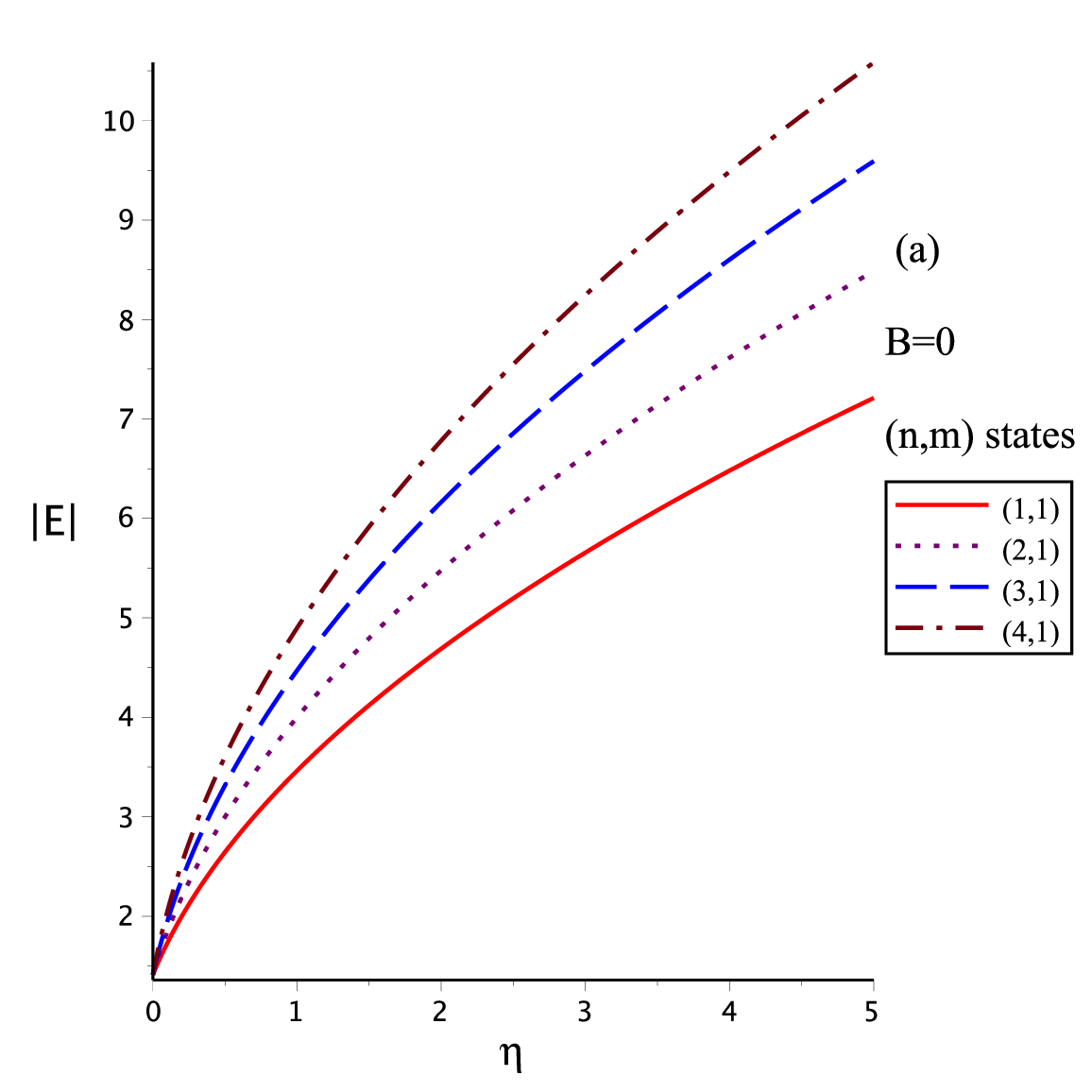}
\includegraphics[width=0.3\textwidth]{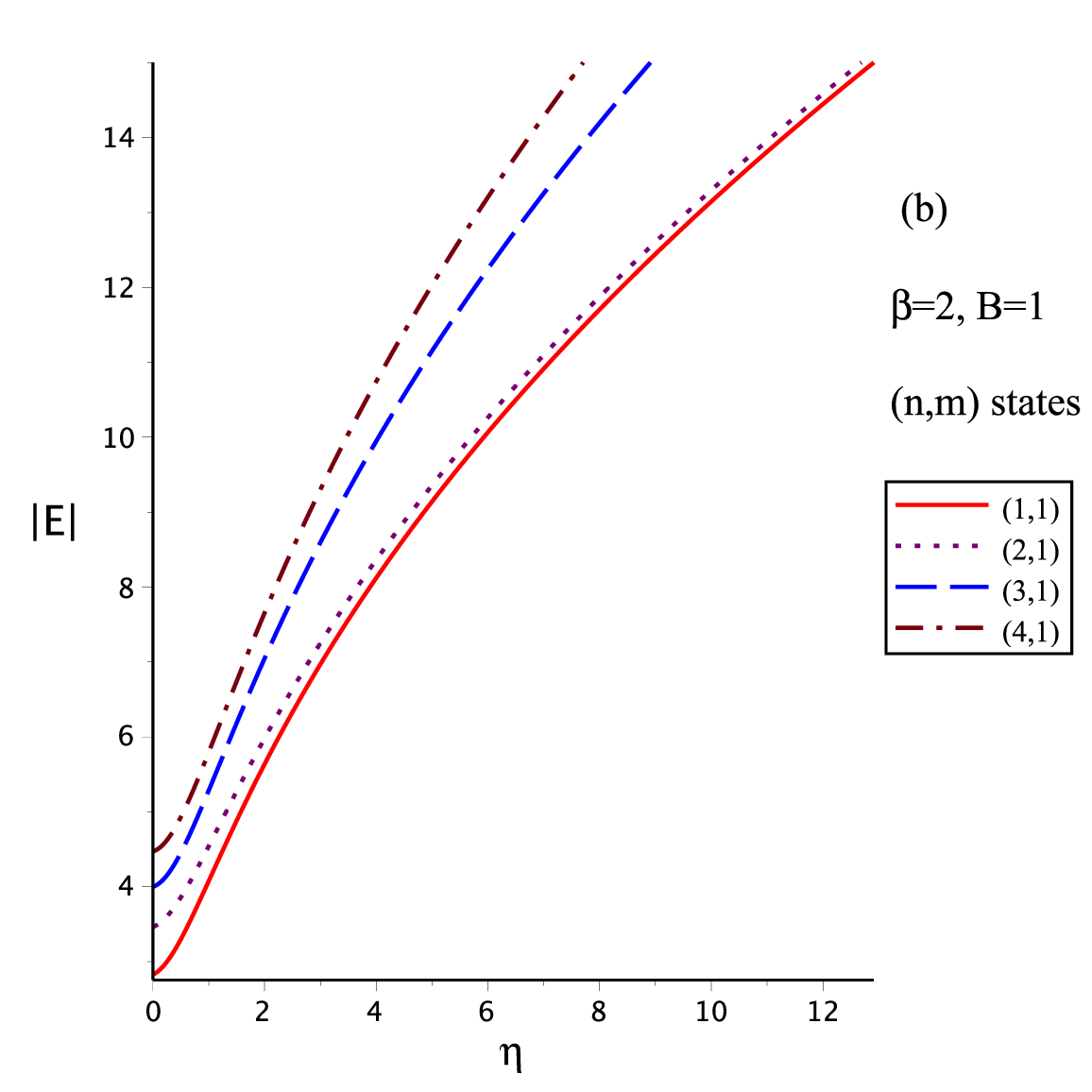}
\includegraphics[width=0.3\textwidth]{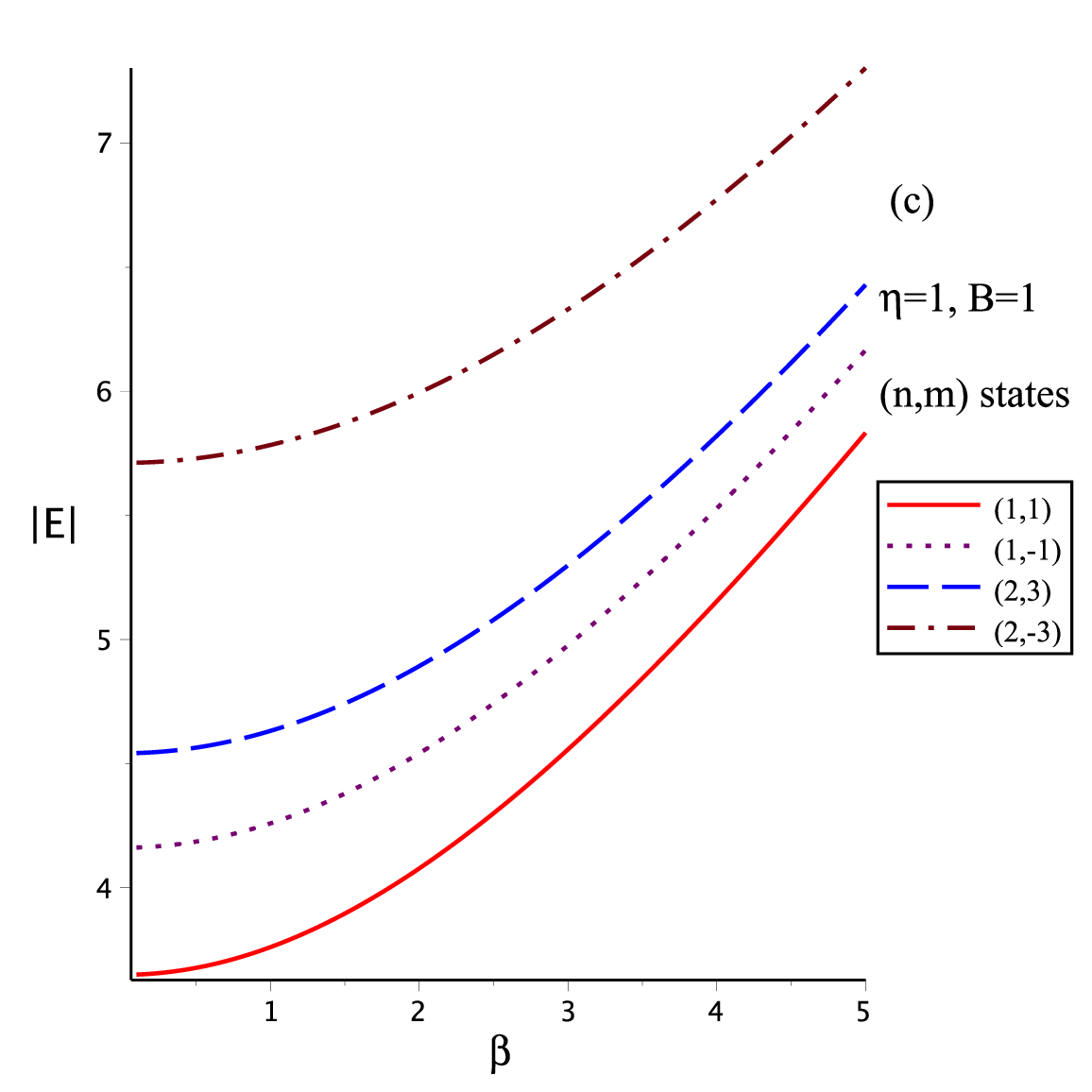}
\caption{\small 
{ The energy levels in Eq. (\ref{4.6}), for the KG-oscillators in a spacetime spiral dislocation II in a magnetic field, with $k=m_{\circ}=e=1$ so that we plot in (a) $|E|$ against $\eta$ with $B=0$ (no magnetic field), (b) $|E|$ against $\eta$ with $B=1$, and (c) $|E|$ against $\beta$ with $B=1$.}}
\label{fig3}
\end{figure}%
We need to truncate the hypergeometric series to a polynomial of order $
n\geq 0$ by requiring $\frac{1}{2}+\nu -\mu =-n$ to eventually imply
\begin{equation}
E_{n,m}=\pm \sqrt{2\Omega \left( 2n+\left\vert m\right\vert +1\right) +k^{2}+m_{\circ }^{2}+2\eta +2\beta ^{2}\tilde{B}\left( \Omega -\tilde{B}\right) -2m\tilde{B}}.
\label{4.6}
\end{equation}
Obviously and interestingly this result shows that the effect of the dislocation parameter $\beta$ on the energy levels, for the KG-oscillators in a spacetime with spiral dislocation II (\ref{2.3})  in a magnetic field, is only feasible for a non-vanishing magnetic field, i.e., $B\neq 0$. Moreover, we observe a Landau-like signature in the result (\ref{4.6}) through the energy terms $2\Omega \left\vert m\right\vert $ and $2m\tilde{B}$. However, one should clearly observe that when the magnetic field is switched off, $\tilde{B}=0$, this result would read
\begin{equation}
    E_{n,m}=\pm \sqrt{2\eta \left( 2n+\left\vert m\right\vert +2\right) +k^{2}+m_{\circ }^{2}}.
\label{4.7}
\end{equation}
and would, consequently, describe the energies for the  KG-Oscillators in a spacetime with spiral dislocation II, (\ref{2.3}), under no magnetic field effect. In this case, nevertheless, the effect of such dislocation II is only observed in the form of the radial part of the wave function in (\ref{4.4}). Yet, such an exact analytical  solvability of  this model problem (\ref{4.2}) allows one to even consider that the KG-oscillators' frequency $\eta=0$ so that the KG-particles in KG-Oscillators in a spacetime with spiral dislocation II, (\ref{2.3}), under only the magnetic field effect admits the form 
\begin{equation}
E_{n,m}=\pm \sqrt{2\tilde{B} \left( 2n+\left\vert m\right\vert +1\right) +k^{2}+m_{\circ }^{2}-2m\tilde{B}}.
\label{4.9}
\end{equation}
In this case, the the magnetic field strength provides the oscillators frequency in terms of $\tilde{B}$ and may very well be classified as KG-magnetic-oscillators. A notion that is manifestly unavoidably introduced by the coexistence of both a spacetime with dislocation II and a magnetic field given by $eA_{\varphi }=\tilde{B}\,r^{2}$.  

In Figure \ref{fig3}, we plot the energy levels of  (\ref{2.6}), for the KG-oscillators in a spacetime spiral dislocation II in a magnetic field, with $k=m_{\circ}=e=1$ so that we plot in \ref{fig3}(a) $|E|$ against $\eta$ with $B=0$ (no magnetic field), \ref{fig3}(b) $|E|$ against $\eta$ with $B=1$, and \ref{fig3}(c) $|E|$ against $\beta$ with $B=1$. We clearly observe that the dislocation parameter's effect on the energy levels is intimately and directly coupled with the magnetic field strength. This is obviously observed through the comparison between the figures in \ref{fig3}(a), \ref{fig3}(b), and \ref{fig3}(c). In \ref{fig1}(a) no magnetic field is presented and the energy levels grow up, as $\eta$ increases, from the same energy point at $\eta=0$. Whereas, in \ref{fig3}(b) and \ref{fig3}(c) we may clearly observe the effect of the dislocation parameter, even at $\eta=0$ for \ref{fig3}(b) and at $\beta=0.1$ for \ref{fig3}(c) (used as the minimum value for $\beta$ in \ref{fig3}(c)), in the form of energy levels splittings.

\section{Summary and discussions}\label{sec4}

In this study, we analyzed the influence of two types of spiral dislocations on the relativistic dynamics of scalar oscillator fields, focusing on both the presence and absence of external magnetic fields. Through our investigation, we derived exact energy spectra and explored how these dislocations and external fields modify the quantum properties of the system.

In the first scenario, in the absence of the magnetic field, where a radial line is twisted into a spiral, we obtained the energy spectrum for the scalar oscillator fields as:
\[E_{nm} = \pm \sqrt{2\eta \left( 2n + |m| + 2 \right) + k^2 + m_{\circ}^2 + \grave{\beta} m^2 },\]
where, \( \grave{\beta} = \beta - \beta^2 \), and the effects of the spacetime background on the energy levels of the KG-oscillator field is observed in the last two terms, under the square root (provided that the range $0<\beta<1$ prevents imaginary energies to be obtained. The effects can also be observed Figure \ref{fig1}. This energy spectrum highlights how the spiral dislocation influences the scalar field’s energy, with the dislocation parameter \( \beta \) modifying the contributions from the quantum number \( m \), without the distinction between $m=+|m|$ and $m=-|m|$ values.  Yet, when an external magnetic field is introduced, the problem becomes conditionally exactly solvable, and in this case the biconfluent Heun solution (\ref{2.12}) suggests that the energies are modified to read
\begin{equation*}
E_{nm}=\pm \sqrt{\left|\sqrt{\grave{\beta}\tilde{B}^{2}+\eta ^{2}}\right| \left( \bar{n}+1/2\right)+\frac{m^{2}\eta ^{2}\grave{\beta}}{\grave{\beta}\tilde{B}^{2}+\eta ^{2}}+k^{2}+m_{\circ }^{2}+\tilde{B}^{2}+2\eta}    \label{2.21}
\end{equation*}
where  \( \bar{n} = 2n + 2|m| + \frac{3}{2} \). This expression reveals the additional influence of the external magnetic field on the energy levels, particularly through the coupling of the magnetic field with the spiral dislocation parameter and the oscillator frequency $\eta$.  However, the conditional exact solvability in this case, offers an exact solution for only a set of the KG-oscillators in spacetime with spiral dislocation I with the dislocation parameter $\beta$ satisfying correlation (\ref{2.17}). In this scenario, effects of the spacetime background and magnetic field on the energy levels of the scalar oscillator field can be seen in the Figure \ref{fig2}.

In the second scenario, where a circle is deformed into a spiral representing an edge dislocation, we obtained the energy spectrum in the absence of a magnetic field as:
\[E_{nm} = \pm \sqrt{2\eta \left( 2n + |m| + 2 \right) + m_{\circ}^2 + k^2}.\]
Interestingly, this result shows that the energy levels of the KG-oscillator field are unaffected by the background spacetime geometry, even though the wave function explicitly depends on the spiral dislocation parameter (see also \cite{2.3,2.4,2.5}). The spiral dislocation influences the structure of the wave functions, but the energy spectrum itself remains unchanged. In the presence of an external magnetic field, however, in this scenario, the energy spectrum becomes:
\[E = \pm \sqrt{2\Omega \left( 2n + |m| + 1 \right) - 2\tilde{B} m + k^2 + m_{\circ}^2 + 2\eta + 2\beta^2 \tilde{B} \left( \Omega - 1 \right)},\]
where \( \Omega = \sqrt{\eta^2 + \tilde{B}^2} \). This spectrum demonstrates how the magnetic field, coupled with the spiral dislocation parameter \( \beta \), significantly modifies the energy levels, particularly through the term involving the magnetic quantum number \( m \).  Notably, moreover, we observe that when $\eta=0$ (i.e., no KG-oscillator) we obtain \[E = \pm \sqrt{2\tilde{B} \left( 2n + |m| + 1 \right) - 2\tilde{B} m + k^2 + m_{\circ}^2 + 2\eta + 2\beta^2 \tilde{B} \left( \Omega - 1 \right)}.\]
This result suggests that the notion of "\textit{KG-magnetic-oscillators}" is unavoidable in the process of considering KG-particles in a spacetime with spiral dislocation II, i.e.,  edge dislocation, in a magnetic field. Figure \ref{fig3} shows the influences of the spacetime background and magnetic field on the energy levels of the scalar oscillator field.

Overall, our results highlight the complex relationship between spiral dislocations and external magnetic fields in shaping the relativistic dynamics of scalar oscillator fields. In both scenarios, the spiral dislocation introduces distinct modifications to the wave function and energy spectra. The presence of an external magnetic field further amplifies these effects, leading to richer energy structures and altered quantum behavior. These findings offer deeper insights into how geometric dislocations and magnetic fields affect relativistic quantum systems, with potential applications in fields such as condensed matter physics, quantum field theory, and curved spacetime studies.

\section*{Data Availability Statement}

Authors can confirm that all relevant data are included in the article and/or its supplementary information files

\end{document}